# Unnecessary risks created by uncontrolled rocket reentries

**Michael Byers**
*Department of Political Science, University of British Columbia, Vancouver, British Columbia, Canada*

**Ewan Wright[1]**
*Interdisciplinary Studies Graduate Program, University of British Columbia, Vancouver, British Columbia, Canada*

**Aaron Boley**
*Department of Physics and Astronomy, University of British Columbia, Vancouver, British Columbia, Canada*

**Cameron Byers**
*Bachelor of Engineering Program, University of Victoria, Victoria, British Columbia, Canada*

## 1. ABSTRACT

In 2020, over 60% of launches to low Earth orbit resulted in one or more rocket bodies being abandoned in orbit and eventually returning to Earth in an uncontrolled manner. When they do so, between 20 and 40% of their mass survives the heat of atmospheric reentry. Many of the surviving pieces are heavy enough to pose serious risks to people, on land, at sea, and in airplanes.

There is no international consensus on the acceptable level of risk from reentering space objects. This is sometimes a point of contention, such as when a 20 tonne Long March 5B core stage made an uncontrolled reentry in May 2021. Some regulators, including the US, France, and ESA, have implemented a 1 in 10,000 acceptable casualty risk (i.e., statistical threat to human life) threshold from reentering space objects. We argue that this threshold ignores the cumulative effect of the rapidly increasing number of rocket launches. It also fails to address low risk, high consequence outcomes, such as a rocket stage crashing into a high-density city or a large passenger aircraft. In the latter case, even a small piece could cause hundreds of casualties. Compounding this, the threshold is frequently ignored or waived when the costs of adherence are deemed excessive.

We analyse the rocket bodies that reentered the atmosphere from 1992 - 2021 and model the associated cumulative casualty expectation. We then extrapolate this trend into the near future (2022 - 2032), modelling the potential risk to the global population from uncontrolled rocket body reentries. We also analyse the population of rocket bodies that are currently in orbit and expected to deorbit soon, and find that the risk distribution is significantly weighted to latitudes close to the equator. This represents a disproportionate burden of casualty risk imposed on the countries of the Global South by major spacefaring countries.

Modern rockets have reignitable engines, allowing controlled reentries into remote areas of the oceans. This, combined with updated mission designs, would eliminate the need for most uncontrolled reentries. Some extra costs would fall on launch providers, including additional fuel for the reentry manoeuvre. Government missions should be able to absorb these extra costs, but they could affect the ability of a commercial launch provider to compete.

Countries in the Global South, whose populations are being put at disproportionate risk by uncontrolled rocket bodies, should demand that major spacefaring countries level the playing field by mandating controlled rocket reentries. This solution, which will have to be coordinated multilaterally, must create meaningful consequences for non-compliance while allowing leeway for those who cannot immediately engage in or afford controlled reentries.

---

[1] Corresponding author: etwright@student.ubc.ca

## 2. BACKGROUND

In July 2022, a Long March 5B rocket was launched, sending a module of the Chinese Tiangong space station into orbit. The 20-tonne core rocket stage was separated from the module and left in orbit uncontrolled, reentering the atmosphere several days later. The breakup was visible from the ground in Malaysia, and significant pieces of debris were found on the ground and in the sea [1]. No injuries or property damage were reported, though debris from a similar launch in May 2020 landed in a village in the Ivory Coast [2]. This was a particularly large rocket body, but launch operators make similar decisions concerning rocket body reentries almost every week, imposing small but still potentially lethal, preventable risks on the world population.

China was criticized for the uncontrolled reentry, including by NASA Administrator Bill Nelson [3]. However, there is no international consensus on the acceptable level of risk, and other states make similar decisions, with rocket bodies reentering the atmosphere uncontrolled about 40 times per year, on average, corresponding to approximately 80-tonnes of rocket body mass reentering annually [4].

Assessing risk on a launch-by-launch basis ignores the larger and growing cumulative risk to people on Earth's surface. Moreover, leaving rocket bodies to reenter the atmosphere uncontrolled is a choice. Rocket bodies can be designed for controlled reentries, directing them, for example, into the ocean, away from populated areas. This can increase the cost for launch operators, but is necessary to achieve a safe, controlled reentry regime.

## 3. THE PROBLEM

Launcher designs vary, but most leave sizeable rocket bodies in orbit. All rockets have 'core' or 'first' stages, which are usually dropped suborbitally, but some, including the Long March 5B core stage, reach orbit. Some rockets also use boosters, which are dropped downrange of launch, usually into the sea. 'Second' or 'upper' stages are often used to carry the payload (one or more satellites) into the required orbit. While some operators, notably SpaceX, land their first stages, their 'upper' stages still reach orbit, and some of them are abandoned there. Indeed, once the payload is released, most operators abandon the rocket body on the same initial orbit as the satellite. If the perigee of that orbit is sufficiently low, gas drag from the upper atmosphere causes the rocket body to lose altitude over time and eventually reenter in an uncontrolled manner; this can occur at any point under the flight path.

About 65% of low Earth orbit (LEO) launches in 2021 resulted in a rocket body being abandoned in orbit [5]. These large space objects pose a collision risk with satellites and other debris, and some rocket bodies explode in orbit, creating significant debris [6]. When an abandoned rocket body reenters the atmosphere, approximately 20-40% of the mass survives reentry and reaches the ground intact [7]. This creates a risk of casualty to people on the ground, in ships and in aircraft. Some recent reentries have illustrated this concern. In addition to the Chinese reentries mentioned above, in March and April 2022 debris from two separate Long March launches landed in India [8]. Nor is this simply a Chinese problem. In 2016, a SpaceX second stage was abandoned in orbit. It reentered one month later over Indonesia, with two refrigerator-sized fuel tanks reaching the ground intact [9].

Some states, such as the US, Europe, France and Japan, impose a 1-in-10,000 risk limit, whereby a rocket body with a projected casualty risk lower than this value is allowed to reenter uncontrolled, and larger risks are supposed to be mitigated through controlled reentries. However, this limit, which is entirely arbitrary, is frequently waived when inconvenient. In the US, where the Orbital Debris Mitigation Standard Practices (ODMSPs) apply to all launches and require that the risk of a casualty from a reentering rocket body is below a 1-in-10,000 threshold [10], the US Air Force waived the ODMSP requirements for 37 of the 66 launches conducted for it between 2011 and 2018, on the basis that it would be too expensive to replace non-compliant rockets with compliant ones [11]. NASA waived the requirements seven times between 2008 and 2018, including for an Atlas V launch in 2015 where the casualty risk was estimated at 1 in 600 [12]. The European Space Agency (ESA) also issues waivers for deviations from their space debris guidelines, though the frequency of this practice is unknown [13].

Due to an increasing global population, remaining below the 1-in-10,000 threshold will be increasingly difficult for any single launch. To illustrate this reality, Fig. 1 shows the casualty area that is needed to keep a reentry below the 1-in-10,000 risk threshold as a function of the given object's orbital inclination and year of reentry. Because rocket bodies are particularly large space objects, with parts including tanks and motors that often survive reentry, they frequently have casualty areas modelled at 10m$^2$ or larger [14]. Without a drastic reduction in the casualty area from rocket bodies, nearly all rocket body reentries will have to be controlled within the next decade to stay under the 1-in-10,000 threshold. Special attention must be paid to rocket bodies that have long residency times in orbit, as they could meet a casualty risk threshold at the time of launch, but exceed it, potentially substantially, at the time of reentry.

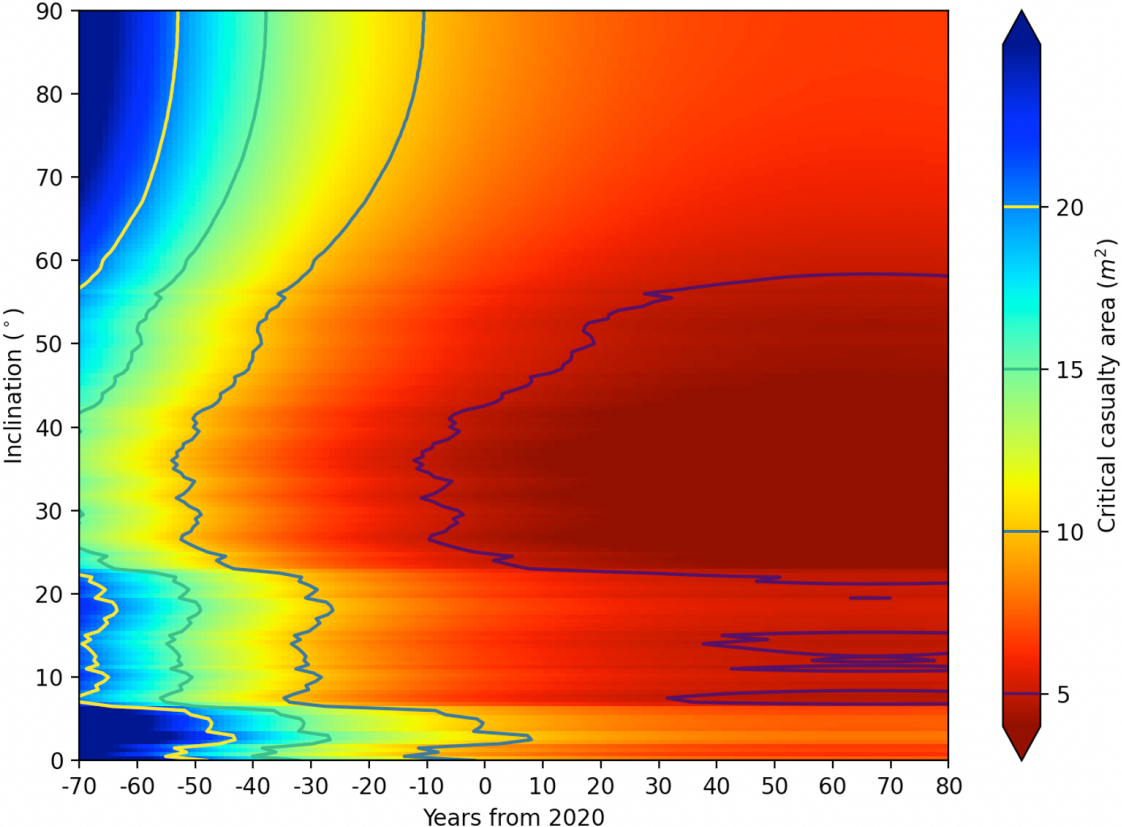

*Fig. 1 Critical casualty area required for a space object reentry to stay under a 1 in 10,000 casualty risk threshold, plotted by object orbital inclination and reentry year. The casualty expectation for each inclination is calculated using the 2020 population [15]. The past critical casualty area is determined back to 1950 (-70 years) using population growth rates from the UN World Population Prospects 2022 dataset [16]. Likewise, the future critical casualty area is calculated using the Median Variant estimations from the same dataset. Based on plot from [14]; independently derived.*

Moreover, the 1-in-10,000 risk limit is not an internationally agreed threshold, and it is estimated that 80% of rocket bodies that reentered from 2010 to 2020 had a casualty expectancy higher than 1-in-10,000 [4]. There are some signs that point toward the development of a new norm, of minimizing risk with or without thresholds. In 2010, the UN Space Debris Mitigation Guidelines recommended that reentering spacecraft not pose 'an undue risk to people or property' [17]. In 2018, the UN Guidelines for the Long-term Sustainability of Outer Space Activities called on national governments to address risks associated with the uncontrolled reentry of space objects [18]. And when the Long March 5B core stage reentered in May 2021, NASA Administrator Bill Nelson said "Spacefaring nations must minimize the risks to people and property on Earth of re-entries of space objects" [19]. But even then, minimizing risk is an ambiguous goal and not a clear standard. There is no binding treaty that addresses rocket body reentries, apart from the 1972 Liability Convention, which stipulates that '[a] launching State shall be absolutely liable to pay compensation for damage caused by its space object on the surface of the earth or to aircraft in flight' [20].

The possibility of liability often induces good behaviour. But apparently, when it comes to rocket bodies, governments have been choosing to bear the slight risk of having to compensate for one or more casualties, rather than to require launch providers to make expensive technological or mission design changes. As in some other areas of government and commercial activity, 'liability risk' is treated as just another cost of doing business [21]. Furthermore, casualty risks are addressed on a launch-by-launch basis, even though the cumulative risk is growing as more and more rockets are launched each year. We argue that this cumulative casualty risk can no longer be ignored. It is time for a globally accepted controlled reentry regime.

## 4. ESTIMATING CASUALTY RISK

In previous work [22], we estimated the casualty risk from reentering rocket bodies using a standard method comparable to that used in [23] and [24]. We used two models to estimate the future risk in different ways, obtaining results that also agreed with a similar and independent analysis conducted by [4]. Sections 4 and 5 are heavily based on our work in [22].

The publicly available satellite catalogue [25] provides data for objects that are currently in orbit, as well as those that have reentered, including rocket bodies. Over the past 30 years (4 May 1992–5 May 2022), more than 1,500 rocket bodies have deorbited [25]. Of these, we estimate that over 70% deorbited in an uncontrolled manner, which (using the 2005 world population for our calculations) corresponds to a casualty expectation of about 0.015 $m^{-2}$. This means that, at face value, if the average rocket body were to cause a casualty area of 10 $m^2$, there was an approximately 14% chance of one or more casualties over this time. Although no such event occurred, or at least was reported, these calculations show that the incurred risk has been far from negligible.

The casualty expectation is calculated as follows. First, a rocket body reentry is taken to be uncontrolled in this analysis if the time span between the rocket's launch date and reentry date is 7 days or longer. Several time spans were tested, and 3-7 days delays yield comparable results. The longer timespan is conservative, corresponding to an estimated 70% of the rocket bodies reentering uncontrolled. Then, since each abandoned rocket body is left at a specific orbital inclination, the probability that an uncontrolled rocket body (or any object) reenters at a given latitude can be expressed through a latitude weighting function. The weight associated with a latitude represents the fraction of time that an object on a fixed inclination spends over the latitude in question. An object on a zero-degree inclination orbit would have a weighting function that is unity at the equator and zero everywhere else, while an object on a 90-degree polar orbit would have a weighting function that is a constant for all latitudes. For all other inclinations, an individual orbit will have a weighting function with peaks at the latitudes close to the value of the orbital inclination, a U-shaped distribution between the peaks, and weights of zero at latitudes higher than the inclination. The weighting function for an individual object is normalized such that the summation of weights over all latitudes is unity, while the weighting function for a population of objects is the sum of the individual functions.

A casualty expectation is determined by taking the product of the weighting function and the population density at a given latitude and summing the result over all latitudes. For reference, Fig. 2C shows the casualty expectation for a single reentering object as a function of its orbital inclination, consistent with previous work [23]. Space objects with an inclination around 30° spend more time over higher population densities and so have a higher casualty expectation. The datasets for the world population for different years are taken from GPWv4 [15]. This basic procedure can also be used to estimate the future risk of uncontrolled rocket body reentries.

## 5. RESULTS

The future rocket body reentry risk can be modelled in several ways; we explore two. First, the long-term risk resulting from the build-up of rocket bodies in orbit can be estimated by looking at which rocket body orbits have a perigee lower than 600 km, with this perigee representing an imperfect but plausible division between rocket bodies that will reenter in the coming decades and those that require much longer timescales. For this cut, there are 651 rocket bodies, with a corresponding casualty expectation of 0.01 $m^{-2}$ (see Fig. 2). Second, we take the trend of rocket body reentries from the past 30 years and apply it to the next 10 years, giving rise to a casualty risk of 0.006 $m^{-2}$ for that period. Both are conservative estimates, as the number of rocket launches is increasing quickly. Assuming again that each reentry spreads lethal debris over a 10$m^2$ area, we conclude that current practices have on order a 10% chance of one or more casualties over a decade.

In the first method (perigee cut), there is no explicit reentry timescale. As such, only the year 2020 world population is used to calculate the corresponding casualty expectation. This method most clearly identifies the consequences of the long-lived on-orbit rocket body population. However, it does not account for the short-lived rocket body population, such as those bodies that reenter within a few weeks after launch. Nor does the method consider world population growth.

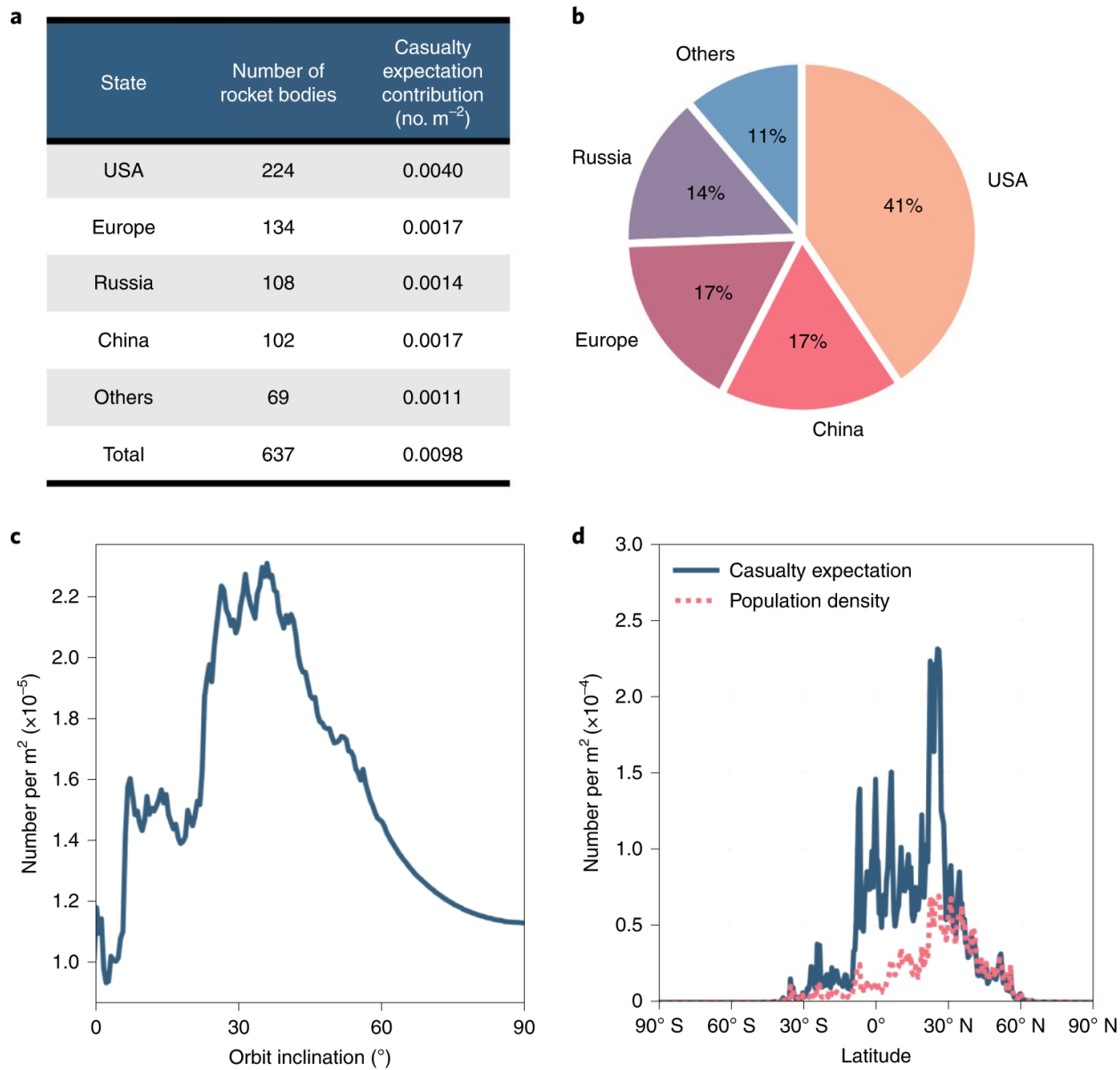

*Fig. 2: Casualty expectations. a, Number of rocket bodies with perigee of <600 km and associated global casualty expectation for spacefaring states with large contributions (Europe treated as a single unit). b, Pie chart of the proportion of the total global casualty expectation contributed by each state. c, Standard casualty expectation as a function of orbital inclination for reentry of a single object and the 2020 global population. d, Casualty expectation of rocket bodies currently in orbit by latitude and 2020 population density.* Casualty expectation is the number of casualties per square metre of casualty area as described in [23]. Casualty area, which is the total area over which debris could cause a casualty for a given reentry, is not modelled. In all panels, only rocket bodies with perigees at or below 600 km are included, based on the satellite catalogue as of 5 May 2022 [25]. This approximates the population of long-lived abandoned rocket bodies that might reasonably be expected to reenter. [22]

In the second method, these shortcomings are addressed, in part, by using the reentry history as a proxy for the future rocket body reentry rate. In this approach, the catalogue is searched for all rocket bodies that have reentered in the past 30 years. Because it is not immediately clear from the catalogue alone which of these reentered uncontrollably, we assume that any rocket body spending more than 7 days in orbit is uncontrolled, as noted above. Finally, the weighting functions for each uncontrolled reentry are averaged over 30 years to arrive at a total average weighting function representative of one year of reentries.

World population growth is modelled as a 1% population increase per year. Assuming no changes to the reentry rate or the distribution of rocket bodies, the resulting total average weighting function is multiplied by the world population density distribution for each year, with the results summed over 10 years. An additional sum over latitude is done to obtain the 10-year casualty risk.

The two methods yield similar results, despite the different approaches. Moreover, the respective weighting functions have a common feature: the largest weights are concentrated near the equator, as shown in Fig. 3.

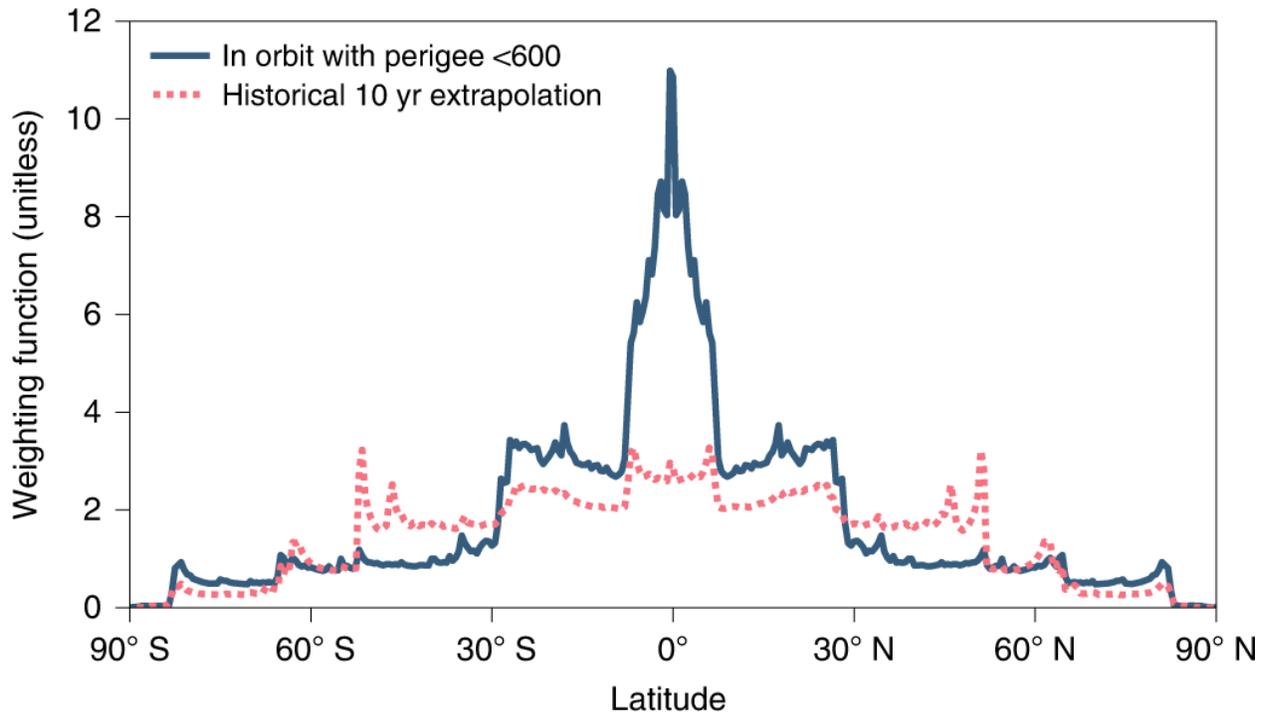

*Fig. 3: Rocket body weighting functions. Each curve is the sum of the rocket bodies' normalized time spent over each latitude. Two models are shown: the sum of all rocket bodies currently in orbit with perigees under 600 km, and a 10-year projection. The latter uses the historical reentries of uncontrolled rocket bodies, from 4 May 1992 to 5 May 2022, to determine an average yearly total weighting function. In this figure, that average is multiplied by ten to show a weighting function for a 10-year period. [22]*

Many of the rocket bodies that lead to uncontrolled reentries are inferred to be associated with launches to geosynchronous orbits, located near the equator. As a result, the cumulative risk from rocket body reentries is significantly higher in the states of the Global South, as compared with the major spacefaring states. The latitudes of Jakarta, Dhaka, Mexico City, Bogotá and Lagos are at least three times as likely as those of Washington, DC, New York, Beijing and Moscow to have a rocket body reenter over them, under one estimate, on the basis of the current rocket body population in orbit (see Fig. 4).

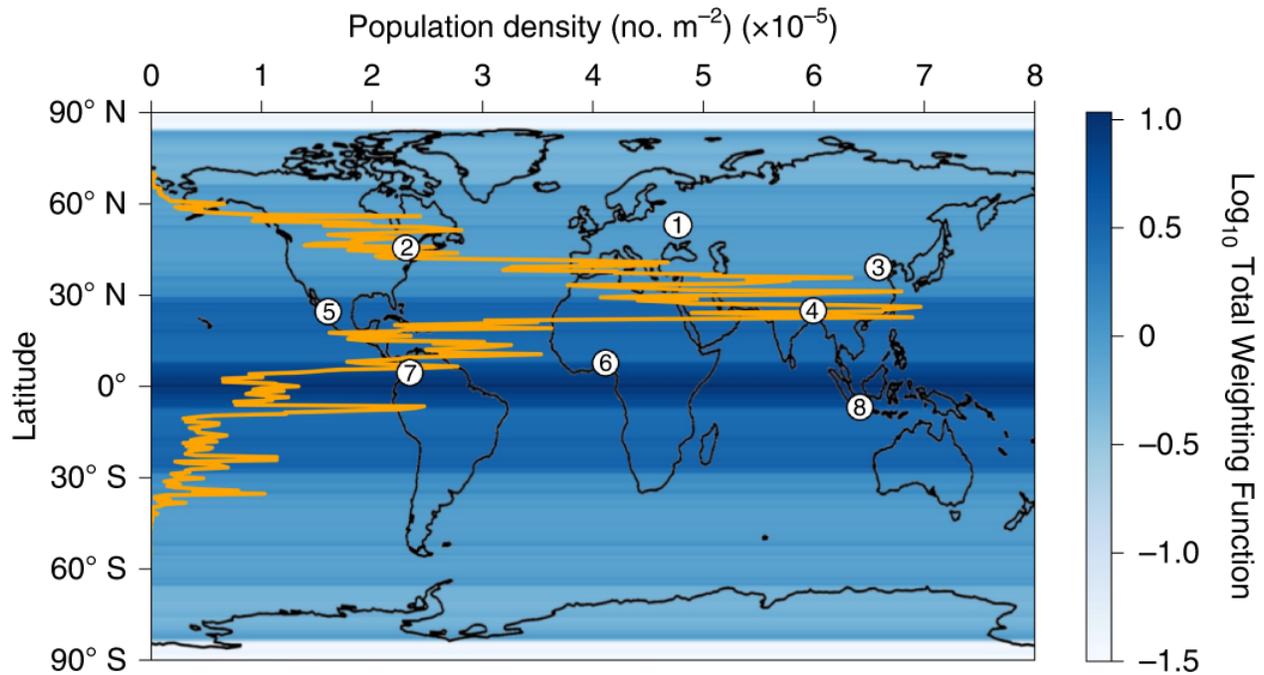

*Fig. 4: World 2020 population density by latitude (plot) and rocket body weighting function (logarithmic heatmap) overlaid on a world map. Some major and high-risk cities are labelled: 1, Moscow; 2, Washington, DC; 3, Beijing; 4, Dhaka; 5, Mexico City; 6, Lagos; 7, Bogotá; 8, Jakarta. The chosen weighting function is for all rocket bodies currently in orbit with perigees less than 600 km in altitude. The outline of the continents is an equirectangular projection, taken from the Python package Cartopy [26]. [22]*

This situation, of risks from activities in the developed world being borne disproportionately by populations in the developing world, is hardly unprecedented. Powerful states often externalize costs and impose them on others, with greenhouse gas emissions being just one example [27]. The disproportionate risk from rocket bodies is further exacerbated by poverty, with buildings in the Global South typically providing a lower degree of protection; according to NASA, approximately 80% of the world's population lives 'unprotected or in lightly sheltered structures providing limited protection against falling debris' [28].

## 6. DISCUSSION

Our casualty risk values agree with similar work that was being conducted concurrently. [4] estimates the total casualty expectancy from 2010 – 2020, considering 613 'large' reentering objects (>1 m$^2$, 68% rocket bodies) as 0.14, corresponding to a 13% chance of a casualty over the period. They explore various fits between masses and casualty areas, with casualty risk estimates for the same population of space objects varying from under 9% to just over 17%.

The future risk estimates are conservative, in part because the number of launches each year is growing. 133 launches were conducted in 2021 [29], a new annual record, and this value is expected to continue growing as the commercial space industry develops and new 'mega-constellations' of satellites are launched.

Casualty expectation modelling also involves some simplifications that are worth exploring. The casualty risk values are dependent on estimates of casualty area, which is the collision cross section between a person and a piece of lethal space debris. This area is modelled as a 0.36 m$^2$ circle, corresponding to a person, tangential to a piece of debris; areas for each debris piece are then summed. However, this ignores scenarios such as a power station or other critical infrastructure being struck, or buildings being caused to collapse, where there could be many direct and indirect casualties. These outcomes are assumed to be balanced out by the protection afforded by some buildings; NASA state that the statistical uncertainty in the casualty expectation formulation "make additional refinement of the equation moot. These uncertainties include true debris size, shape, count, and damage energy, and even the ratio of 'average human' time spent in seated/prone/standing positions" [28].

The relationship between the mass of the re-entering stage and the resulting casualty area is not clear, but some trends have been observed. As noted above, approximately 20-40% of the object will survive reentry, on average [7]. Systems with high melting point materials, or dense materials, will typically survive reentry, as will large tanks. About 150 pieces of space debris have been found across more than 67 impact zones; nearly 40% of these pieces were pressure vessels [30]. As it is unlikely that all, if not most, pieces of a reentering space object will be recovered for analysis, demise of space objects is typically modelled computationally. Due to the 0.36 m$^2$ human size included in every casualty area, the driving factor behind the total casualty area is the number of pieces that reach the ground; a cloud of golf ball-like objects is modelled to be more damaging than one large piece, although the latter could destroy an apartment block housing many people.

Furthermore, high consequence events, such as an aircraft being hit by debris from an uncontrolled rocket body reentry, are not modelled. If even a small piece of debris struck an aircraft it could cause tens or hundreds of casualties, far exceeding our estimates. According to [31], only a 300 g piece of debris could catastrophically damage an airliner in flight. But [32] estimates that pieces of metal of 1 – 3.5 g could lethally damage windshields or engines. A 2008 analysis by [33] assumed 100 large object reentries per year, resulting in a probability that an aircraft is struck each year of 7.69 x10$^{-5}$. This only considered US aircraft but estimated that aircraft worldwide have an annual risk of approximately 3 x10$^{-4}$ impacts per year. This risk is still small but again, the cumulative risk is growing, and the consequences of such an event would be large. Even with the 2008 space debris population, it is more likely that an aircraft will be struck by reentering space debris than by a meteorite [33].

Our research used a 10 m$^2$ casualty area that is likely to be an underestimate. Reference [15] used models that estimate some rocket body casualty areas as large as 40 m$^2$, and several 1-tonne rocket bodies with casualty areas of 20 m$^2$. Using data from [34], we analyzed 523 rocket bodies with perigees under 600 km and masses greater than 100 kg, using a fit outlined in [14], leading to a mean casualty area of 23 m$^2$.

There are other simplifications. In the satellite catalogue, some rocket body-like objects (namely, separate apogee motors) are included as payloads; arguably these are rocket bodies. Objects are more likely to deorbit near the equator due to the oblate shape of Earth [35], but this factor is not included in models. And the population will not grow uniformly – countries in the Global South typically have larger growth rates.

While incorporating these factors would increase the detail of the model, the order of magnitude of the result would be comparable. Our main finding is that these risks, when considered cumulatively, are no longer negligible, and they are growing. They are also completely preventable.

## 7. A COLLECTIVE ACTION PROBLEM

A risk to a single life might be seen as a cost of doing business. But the potential ramifications of casualties from an uncontrolled rocket body reentry could be much larger than generally assumed. For example, a mass casualty event in one country caused by a reentering rocket body from another could create geo-political tension, similar to when an aircraft is accidentally shot down.

Yet it is no longer necessary to abandon rocket bodies in orbit, leaving them to reenter the atmosphere in an uncontrolled way. With a bit more fuel and engines that can reignite, controlled reentries can be achieved [36]. Of course, any controlled reentry regime will result in extra costs, and while government missions should be able to absorb these costs, they could affect the ability of a commercial launch provider to compete. National governments could raise the standards applicable to launches from their territory or by companies incorporated there, but individual governments might have competing incentives, such as reducing their own costs or growing a globally competitive domestic space industry.

Uncontrolled rocket body reentries thus constitute a collective action problem; solutions exist, but every launching state must adopt them. The challenge is to prevent 'free-riders', whether in the form of states that seek to retain a short-term competitive advantage by continuing to allow uncontrolled reentries, or even states that seek to attract foreign launch operators by so doing.

We have been here before. The 1970s saw a growing risk to oceans and coastlines from oil spills as well as efforts, nationally and internationally, to adopt a requirement for 'double hulls' on tankers. The shipping industry, concerned about increased costs, was able to stymie these efforts—until 1989, when the Exxon Valdez spilled roughly 11 million gallons of oil into Alaska's Prince William Sound. Media coverage of the accident made the issue of oil spills a matter of public concern, and after the National Transportation Safety Board concluded that a double hull would have substantially reduced if not eliminated the spill [37], the US government required all new tankers calling at US ports to have double hulls [38]. This unilateral move then prompted the International Maritime Organization to amend the International Convention for the Prevention of Pollution from Ships (MARPOL Convention) in 1992 to require double hulls on new tankers and, through further amendments in 2001 and 2003, to accelerate the retirement of single-hulled tankers.

The 1992 amendments to the MARPOL Convention have since been ratified by 150 states (including the US, Liberia and Panama), representing 98% of the world's shipping tonnage. This precedent, of oil spills and the double-hull requirement, is especially relevant for uncontrolled rocket body reentries because it concerns transportation safety in an area beyond national jurisdiction, with oil spills posing risks for all coastal states.

Now, we need a widely accepted controlled reentry regime for rocket bodies. Negotiations on this matter will take several years, and probably longer, which is why they will need to start soon. In the meantime, responsible states should do everything they can to minimize the risks—including by unilaterally declaring their commitment to transition away from uncontrolled reentries.